# Optimal Dispatch Strategy for a Multi-microgrid Cooperative Alliance Using a Two-Stage Pricing Mechanism

Yonghui Nie, *Member, IEEE*, Zhi Li, Jie Zhang, Lei Gao, Yang Li, *Senior Member, IEEE*, and Hengyu Zhou

*Abstract*— To coordinate resources among multi-level stakeholders and enhance the integration of electric vehicles (EVs) into multi-microgrids, this study proposes an optimal dispatch strategy within a multi-microgrid cooperative alliance using a nuanced two-stage pricing mechanism. Initially, the strategy assesses electric energy interactions between microgrids and distribution networks to establish a foundation for collaborative scheduling. The two-stage pricing mechanism initiates with a leader-follower game, wherein the microgrid operator acts as the leader and users as followers. Subsequently, it adjusts EV tariffs based on the game's equilibrium, taking into account factors such as battery degradation and travel needs to optimize EVs' electricity consumption. Furthermore, a bi-level optimization model refines power interactions and pricing strategies across the network, significantly enhancing demand response capabilities and economic outcomes. Simulation results demonstrate that this strategy not only increases renewable energy consumption but also reduces energy costs, thereby improving the overall efficiency and sustainability of the system.

## I. INTRODUCTION

In recent years, the gradual depletion of non-renewable energy sources and air pollution caused by combustion have emerged as significant obstacles to global economic development [1]. The development and application of distributed energy sources, such as wind power (WP) and photovoltaic (PV), offer viable solutions to these challenges [2]. A new-energy microgrid (NEMG), an important component of distributed energy system [3], enhances the operational efficiency and utilization of renewable energy by optimizing interaction among distributed power sources, loads, and energy storage. As NEMGs gain broader access to the distribution network, multiple geographically adjacent microgrids can be interconnected to form a multi-microgrid (MMG) system [4]. This system facilitates power exchanges within and between microgrids and distribution networks, thereby enhancing both the flexibility and reliability of the grid. Compared to a single microgrid, MMG systems substantially improve the integration of distributed energy, reduce power fluctuations, and promote the economic viability of NEMGs [5].

Electric vehicles (EVs), as mobile loads with potential for energy savings and low emissions, feature bidirectional energy storage capabilities [6]. This characteristic enables them to participate in MMG scheduling as both producers and consumers. Utilizing vehicle-to-grid (V2G) technology, EVs help reduce electricity costs, smooth out fluctuations in distributed energy, and provide the MMG with auxiliary services such as energy storage, peak shaving, and frequency regulation [7]. However, the integration of EVs also complicates interactions among MMG stakeholders, presenting significant challenges for system operation and management. Therefore, examining the electric energy transaction mechanisms in detail becomes crucial after EVs are integrated into the MMG, ensuring its economical operation.

With the increasing penetration of renewable energy in microgrids, the impact of power fluctuation of this type of energy on system security and stability will be exacerbated [8]. A cooperative alliance [9], as one of the grouping operation methods, can form a community of interest among multiple microgrids to guarantee the economic and stable operation of the system. Reference [10] improved the economic efficiency of the MMG cooperative alliance system by signing agreements between the subnets, thereby equating all microgrids into an alliance as a whole. Reference [11] established an alliance scheduling model for a MMG system by considering cogeneration, storage, and distribution of energy, which reduces the operating cost of microgrids as well as the total cost of scheduling in the MMG cooperative alliance. Although the previous studies have improved the economic efficiency of MMGs through cooperative alliances, they have not considered the impact of the electricity consumption behavior generated by the demand-side interest interactions on the MMG system.

This work was supported by the Science and Technology Development Plan Project of Jilin Province under Grant 20230101216JC and 20240304157SF. Paper no. TSTE-00122-2024. *(Corresponding author: Yonghui Nie, Yang Li.)*

Yonghui Nie is with the Academic Administration Office, Northeast Electric Power University, Jilin 132012, China (e-mail: nieyonghui@neepu.edu.cn).

Zhi Li and Yang Li are with the Department of Electrical Engineering, Northeast Electric Power University, Jilin 132012, China (e-mail: 2202200330@neepu.edu.cn; liyang@neepu.edu.cn).

Jie Zhang is with the State Grid Chengwu Electric Power Supply Company, Chengwu, 274200, China (e-mail: 313826289@qq.com).

Lei Gao is with China Electric Power Research Institute, Beijing, 100192, China (e-mail: gaolei@epri.sgcc.com.cn).

Hengyu Zhou is with the State Grid Yingkou Electric Power Supply Company, Yingkou, 115000, China (e-mail: 293528896@qq.com).

For a MMG system, demand response (DR) is a crucial strategy for balancing energy supply and demand during the cost optimization process. DR involves the short-term adjustment of electricity consumption by power users in response to price changes, thereby smoothing peaks and enhancing the utilization of distributed energy resources. Reference [12] established a MMG system model that considers the randomness and diversity of demand-side electricity consumption. By combining price incentives and scheduling potential to guide load participation in response, the operating cost of household MMG systems is effectively reduced. Reference [13] incorporated WP and PV into a MMG system through a multi-agent system and implemented price-based DR on the load side, ensuring the economic operation of the MMG system. These studies only consider the economic demand on the demand side as a constraint to bind the interests of the microgrid system when studying comprehensive energy trading, ignoring the interest interaction relationship between the NEMG and power users. Reference [14] incorporated EVs into the demand response scope and improved the profits of new-energy suppliers, comprehensive energy service providers, and EV users by constructing an optimization scheduling model based on multi-party interest games. Reference [15] established a multi-master, multi-slave MMG scheduling strategy with each microgrid within a MMG system as a leader and the users within the microgrids as followers.

Despite these advances, existing literature does not adequately address the precision required in demand response strategies for MMG systems when both user loads and EVs are involved in scheduling simultaneously. Table I illustrates the distinctions between the model proposed in this study and the most relevant research in the field, identifying several critical issues in energy supply and trading within MMG systems that incorporate EVs: (1) Most current research treats the DR of EVs and users independently, neglecting the intricate coupling relationship and the sequential interactions when both are engaged in DR simultaneously. (2) The scheduling strategies frequently overlook the variability of customer loads and the intermittency of renewable energy outputs, such as wind and solar power, particularly concerning the integration of EVs. (3) The optimization of MMG scheduling rarely considers the complex interests of multiple stakeholders participating in DR at the same time, which can lead to suboptimal energy management and economic inefficiencies.

To address these issues, this study proposes an optimal dispatch strategy for a MMG cooperative alliance based on a two-stage pricing mechanism. The main contributions are as follows:

Table I
Comparison of the proposed model with the most relevant studies

| Reference | Stakeholders | | Scheduling modeling method | Demand response | | Consider the ride satisfaction of EV owners | Renewable Uncertainties | | |
| --- | --- | --- | --- | --- | --- | --- | --- | --- | --- |
| | Upper-level | Lower-level | | User variable loads | EVs | | User loads | WT | PV |
| [7] | Residential energy systems | EV owners + energy storage | Two-layer optimization modeling | × | √ | √ | × | × | × |
| [9] | Multi-microgrid operator | Users + Shared energy storage | Multi-microgrid cooperation alliance + Stackelberg game | √ | × | × | × | × | × |
| [11] | Multi-microgrid operator | Users+Energy Storage | Stackelberg game | √ | × | × | × | × | × |
| [13] | Distributed energy operator | EV owners | Stackelberg game | × | √ | √ | × | × | × |
| [14] | Multi-microgrid operator | Users | Two-layer optimization modeling | √ | × | × | × | √ | √ |
| [26] | Multi-microgrid operator | EV owners | Two-layer optimization modeling | × | √ | × | √ | √ | √ |
| This paper | Multi-microgrid operator | Users + EV owners | Multi-microgrid cooperation alliance + TPM | √ | √ | √ | √ | √ | √ |

1) This study proposes a two-stage pricing mechanism (TPM) that utilizes time-of-use (TOU) tariffs and dynamic tariffs to achieve hierarchical scheduling. This approach meticulously tunes the participation of both general users and EV owners in demand response activities, ensuring that their diverse interests are harmoniously aligned, which is crucial for maintaining system balance and user satisfaction.

2) Our research pioneers the concept of cooperative alliances among microgrids by proposing a dispatch model that enhances the collective management of multiple microgrids. The MMGs are organized into a community of interest, facilitating the flexible consumption of regional renewable energy to meet the power demands of different microgrids. The Shapley algorithm is employed to coordinate the distribution of benefits among microgrids, thereby promoting economic efficiency and system stability.

3) An arithmetic example is analyzed to demonstrate the substantial benefits of the two-stage pricing mechanism, particularly in enhancing supply management and demand-side energy utilization within the MMG system. This analysis reveals potential improvements for the pricing mechanism, especially in systems that incorporate EVs, thereby offering critical insights into optimizing energy distribution and usage.

## II. MULTI-MICROGRID SYSTEMS FOR INTEGRATING DISTRIBUTED ENERGY AND ELECTRIC VEHICLES

### A. MMG system structure

A MMG system comprises multiple NEMGs, each consisting of EVs, charging stations, smart meters, wind turbines (WT), photovoltaic panels, and basic electric loads, such as refrigerators and dishwashers. The MMG structure is illustrated in the figure below:

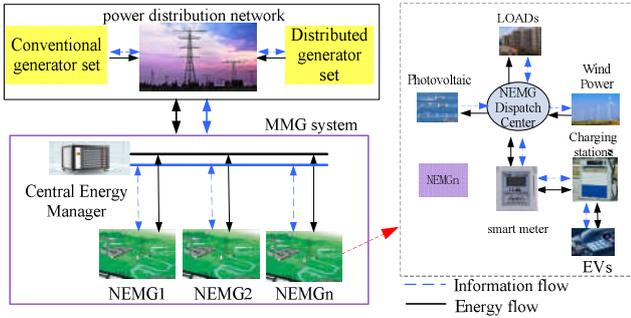

**Fig. 1.** Schematic of a multi-microgrid system.

In the grid-connection method for the MMG, this study implemented a centralized topology where each NEMG is uniformly managed by a central energy manager [16]. Within each NEMG, the smart meter serves as the foundational component, enabling metering, two-way communication, and device control. When an electric vehicle is connected to a charging station, it not only receives demand information from the station owner but also transmits commands to the smart meter. This configuration allows the smart meter to precisely manage the vehicle's charging and discharging processes. The detailed process is depicted in Fig. 2.

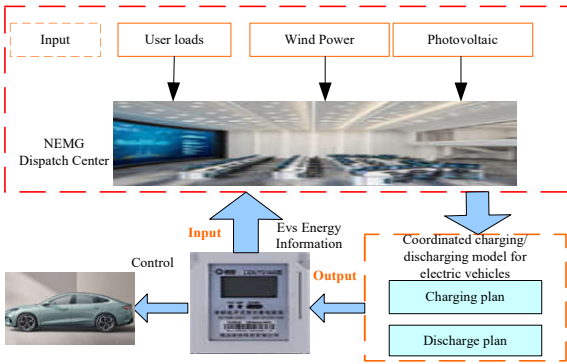

**Fig. 2.** Flow chart of charging control for EVs

### B. Probability distribution and uncertainty reduction of WT and PV output

The main factors affecting the wind power output are wind field location, environment, and other objective factors. Among them, the wind speed is one of the main factors and, according to statistics, the wind speed generally obeys the WeiBull distribution. The probability density function is

$$f(v) = \frac{m}{g}(\frac{v}{g})^{k-1} \exp\left[-(\frac{v}{g})^m\right], \quad (1)$$

where $v$ is the wind speed, $m$ is the shape parameter, and $g$ is the scale parameter.

The expression for fan output is as follows:

$$P^{WT} = \begin{cases} 0 & v \leq v_{ci} \\ k_{w1}v^3 - k_{w2}P_{WT\_rated} & v_{ci} < v \leq v_r \\ P_N & v_r < v \leq v_{co} \\ 0 & v > v_{co} \end{cases} \quad (2)$$

where $v_{ci}$ is the cut-in wind speed, $v_{co}$ is the cut-out wind speed, $v_r$ is the rated wind speed, and $P_N$ is the rated output power of the wind turbine.

The coefficients of the function curve corresponding to the rise in fan output are

$$k_{w1} = P_N / \left(v_r^3 - v_{ci}^3\right), \quad (3)$$

$$k_{w2} = v_{ci}^3 / \left(v_r^3 - v_{ci}^3\right). \quad (4)$$

Photovoltaic intensity and temperature are the main factors affecting PV power generation and their output voltammetric characteristics [17] are expressed in (5):

$$I = I_{SC}\left\{1 - C_1\left[\exp\left(\frac{U - \Delta U}{C_2 U_{OC}}\right) - 1\right]\right\} + \Delta I. \quad (5)$$

The relevant parameters are as follows:

$$\begin{cases} C_1 = (1 - I_m / I_{sc})\exp(-U_m / C_2 U_{oc}) \\ C_2 = (U_m / U_{oc} - 1) \cdot \dfrac{1}{\ln(1 - I_m / I_{oc})} \\ \Delta I = a\Delta T A / A_{ref} + (A / A_{ref} - 1)I_{sc} \Delta U = -b\Delta T - R_s \Delta I \\ \Delta T = T - T_{ref} \\ T = T_a - K_c A \end{cases} \quad (6)$$

where $I_{sc}$ is the short-circuit current of the PV cell; $U_{oc}$ is the open-circuit voltage of the PV cell; $U_m$ and $I_m$ are the voltage and current corresponding to the maximum power point, respectively; $A$ and $A_{ref}$ are the intensity of arbitrary solar radiation and its reference value, respectively; $T_s$ is the series resistance of the PV array; $T$, $T_{ref}$, and $T_a$ are the temperature of the solar panel, the reference value of the temperature of the PV cell, and the ambient temperature, respectively; $K_c$ is the temperature coefficient of the PV panel.

### C. Scenario generation and reduction

To reduce the uncertainty risk caused by energy sources, we adopted Latin hypercube sampling (LHS) [18-19] for sampling the relevant parameters of the WT, PV, and Load and generate the output scenarios as follows:

$$P_{s,t}^{\text{ID}} = P_t^{\text{DA}}\left(1+\tau(T_t-\lambda)\right) \quad (7)$$

where $P_{s,t}^{\text{ID}}$ is the predicted value, $P_t^{\text{DA}}$ is the historical power, $\tau$ is the prediction error factor, $T_t$ represents a random number generated by the distribution function obeyed by the distributed energy sources and loads, and $\lambda$ is the random distribution correction factor.

According to previous studies, the prediction error of the wind power output generally obeys a beta distribution [20], and the prediction errors of the PV and load output are characterized by a normal distribution [21]. The specific probability distribution functions are given by Eqs. (8) and (9), respectively.

$$f(K) = \frac{1}{\sigma\sqrt{2\pi}} e^{-\frac{(\omega-\mu)^2}{2\sigma^2}} \quad (8)$$

$$f(K) = N_d K_{\alpha-1,t}(1-K_t)^{\beta-1}, \quad (9)$$

where $\mu$ and $\sigma$ represent the expectation and variance of the normal distribution, respectively; $\alpha$ and $\beta$ are the shape and scale parameters of the beta distribution, respectively; and $N_d$ is the normalization factor. Specifically, $\mu$ and $\sigma$ are set as 0.5 and 0.33, respectively, $\alpha$ and $\beta$ are set as 2.5, and $\gamma$ is set as 0.5.

To filter out scenarios with planning significance, the number of scenarios must be reduced. In this study, the K-means clustering algorithm was utilized to reduce the generated outgoing scenarios. Its main steps are as follows:

Step 1: Calculate the closest scene $x_j$ to scene $x_i$ among all generated scenes.

$$D_i = \min\left(\lambda_i\, d(w_i,w_j)\right), j=1,2,\cdots,n_{sc}\text{ and }j\neq i \quad (10)$$

where $D_i$ represents the probabilistic distance between scene $w_j$ and scene $w_i$; $d(w_i,w_j)$ represents the Euclidean distance between scene $w_j$ and scene $w_i$; and $n_{sc}$ is the number of initially generated scenes.

Step 2: Identify scenes to be deleted $x_i$。

$$D_{\min} = \min_{i=1,2,L,n_{sc}}(\lambda_i D_i), \quad (11)$$

where $D_{\min}$ is the probabilistic distance to the closest scene $w_i$.

Step 3: Delete the scene $x_i$ identified in Step 2 and add the probability of $w_i$ to the probability of $w_j$ for the closest sample to ensure that the sum of the probabilities for all scenarios is always unity. The probability of $w_j$ after deleting scene $w_i$ is given by Eq. (12).

$$\lambda_j' = \lambda_j + \lambda_i \quad (12)$$

Step 4: Repeat the above steps until the number of remaining scenes reaches a set value.

### D. Analysis of charging load characteristics of electric vehicles

Currently, EVs are categorized into two main types: private and public. Public EVs operate on a fixed schedule, meaning their charging times are generally unaffected by fluctuations in electricity prices. Consequently, they maintain a consistent load during morning and evening peak periods and lack dispatchability. In contrast, the charging and discharging behavior of household EVs is significantly influenced by electricity prices, granting these owners greater flexibility in scheduling their charging times [22]. Consequently, this study utilizes a household EV charging load characteristic model to simulate various strategic scenarios.

According to the investigation results of the US Department of Transportation on EVs, the main factors affecting the charging and discharging load of EVs are the end time of daily driving, return time, and daily driving distance, which are approximately normal distribution functions.

The return time of EVs obeys the following probability density function [23]:

$$f_s(x) = \begin{cases} \frac{1}{\sqrt{2\pi}\sigma_s}\exp\left[-\frac{(x+24-\mu_s)^2}{2\sigma_s^2}\right] \\ 0 < x \leq \mu_s - 12 \\ \frac{1}{\sqrt{2\pi}\sigma_s}\exp\left[-\frac{(x-\mu_s)^2}{2\sigma_s^2}\right] \\ \mu_s - 12 < x \leq 24 \end{cases}, \quad (13)$$

where $\mu_s = 17.47$ and $\sigma_s = 3.41$.

The probability density function for the first travel time is given in [24]:

$$f_e(x) = \begin{cases} \frac{1}{\sqrt{2\pi}\sigma_e}\exp\left[-\frac{(x-\mu_e)^2}{2\sigma_e^2}\right] \\ 0 < x \leq \mu_e + 12 \\ \frac{1}{\sqrt{2\pi}\sigma_e}\exp\left[-\frac{(x-24-\mu_e)^2}{2\sigma_e^2}\right] \\ \mu_e + 12 < x \leq 24 \end{cases}, \quad (14)$$

where $\mu_e = 8.92$ and $\sigma_e = 3.24$.

The probability density function of daily mileage of EVs is provided in [25]:

$$f_m(x) = \frac{1}{\sqrt{2\pi}\sigma_m x}\exp\left[-\frac{(\ln x - \mu_m)^2}{2\sigma_m^2}\right], \quad (15)$$

where $\mu_m = 2.98$ and $\sigma_m = 1.14$.

## III. BI-LEVEL OPTIMIZATION MODEL AND SOLUTION METHOD

To realize the MMG system and the demand side optimization, a bi-level optimization model is introduced in this paper [26]. At the upper level, the model is designed to minimize the daily operating costs of the MMG. It utilizes power planning data from the demand side of the sub-microgrids, provided by the lower level, to configure the interactive power necessary to achieve balance across the NEMGs. The lower level employs a two-stage pricing mechanism that facilitates hierarchical optimization of electricity prices for users and EVs, along with planning their electricity consumption. The structure diagram of this bi-level model is shown in Fig. 3.

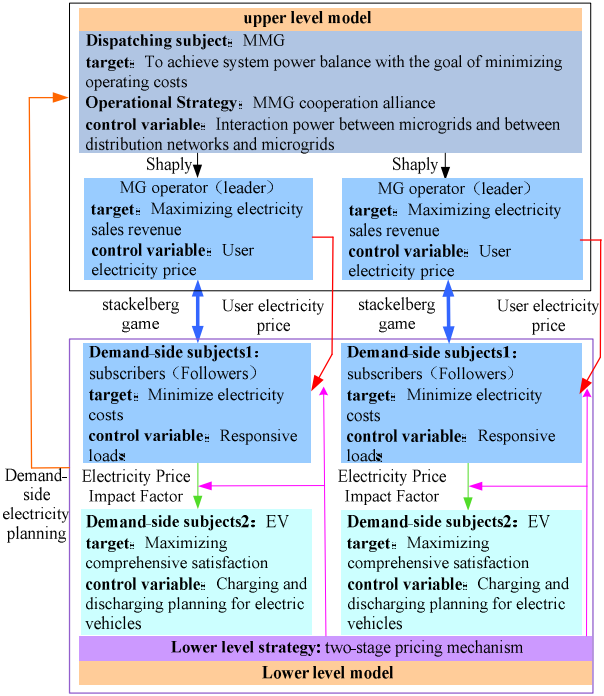

**Fig. 3.** Structure diagram of double-layer optimization model.

## A. Upper-level optimization model

To effectively coordinate NEMGs with different energy characteristics, reduce the impact of power fluctuations of the new-energy sources on system security and stability, guarantee the power demand of microgrids, and minimize the system operation cost, this study adopted the idea of cooperative alliance in the upper level and established a cooperative alliance model for MMGs, as shown in Fig. 4.

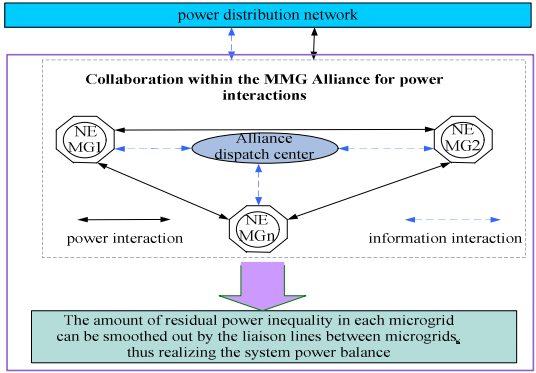

**Fig. 4.** MMG cooperation alliance framework.

In the MMG cooperative alliance system, the alliance scheduling center initially coordinates the power interactions among according to the optimization results. If the internal power supply of the alliance is insufficient to meet the NEMG's demand, additional power is sourced from the distribution grid. The trading methods employed in the MMG cooperative alliance strategy developed in this study consist of three key components:

Firstly, electricity trading between the MMG and power distribution network; secondly, electricity trading between sub-microgrids, each NEMG can sell the excess power or buy power from other microgrids based on the pre-established agreed tariff. Finally, through hierarchical electricity price incentives, the demand side is guided to further absorb or supplement the power of the NEMG.

After obtaining the total operating cost of the system, the Shapley value method is used to allocate the overall operating cost of the alliance, thereby achieving an increase in the benefits of each participant.

*1) Objective Function:* In a MMG cooperative alliance, the individual participants have to implement the alliance decisions. Therefore, in the calculation, all NEMGs can be equated as a whole, and the benefit claim of the MMG can be expressed as the minimized operating cost:

$$\min\ F_{MMG} = F_1 + F_2 + F_3 + F_4. \qquad (16)$$

The daily operating cost of the MMG consists of the transaction cost $F_1$ between the alliance and the distribution grid, the transaction cost $F_2$ within the alliance, the generation cost $F_3$ of the MMG, and the operation and maintenance cost $F_4$ of the public contact line.

$$F_1 = \sum_{t=1}^{24}(C'_d P^{sell}_{mmg,t} - C_d P^{buy}_{mmg,t}), \qquad (17)$$

where $P^{buy}_{mmg,t}$ and $P^{sell}_{mmg,t}$ denote the power purchased/sold by the MMG to the power distribution network; $C_d$ and $C'_d$ are the price of power purchased and sold to the distribution network.

$$F_2 = \sum_{k=1}^{N}\sum_{t=1}^{24}(C_m P^{buy}_{k,t} - C_m P^{sell}_{k,t}), \qquad (18)$$

where $C_m$ represents the power purchase/sale price between NEMGs, $P^{buy}_{k,t}$ and $P^{sels}_{k,t}$ are the power purchase/sale to other microgrids, and $N$ is the total number of microgrids.

$$F_3 = \sum_{k=1}^{N}\sum_{t=1}^{24}(C_{PV} P^{PV}_{k,t} + C_{WT} P^{WT}_{k,t}), \qquad (19)$$

where $C_{PV}$ and $C_{WT}$ represent the cost of PV and wind power generation; $P^{PV}_{k,t}$ and $P^{WT}_{k,t}$ are the PV and WT output power of the NEMG.

$$F_4 = \gamma_L \sum_{t=1}^{24}(P^{sell}_{mmg,t} + P^{buy}_{mmg,t}), \qquad (20)$$

where $\lambda_L$ is the operation and maintenance cost of 1 kw.h of electricity delivered by the contact line.

After determining the overall operating costs, individual microgrid cost-sharing solutions are developed using the Shapley value method.

A total of $(|S|-1)!$ NEMGs participate in coalitional cooperation kinds of ordering, where |S| is the number of partial sub-microgrids, and the remaining N-|S| NEMGs are ordered in (N-|S|)! kinds. The different ordering combinations of participating NEMGs divided by the randomized ordering combinations of N microgrids is the weight of the microgrid's share of benefits in the whole coalition:

$$\frac{(|S|-1)!(N-|S|)!}{N!}. \quad (21)$$

The benefits that the NEMG participating in different coalitions creates for itself and the coalition are expressed in (22):

$$v(S) - v(S \setminus \{i\}). \quad (22)$$

Thus, the microgrid's share of the overall benefits can be obtained as shown in (23).

$$\varphi_i(v) = \sum_{S \subset M} \frac{(|S|-1)!(N-|S|)!}{N!}(v(S) - v(S \setminus \{i\})), \quad (23)$$

where $M = \{1, 2, \cdots, N\}$; $v(S)$ is the revenue gained by the coalition when the NEMG participates in the coalition cooperation; $v(S\setminus\{i\})$ is the revenue gained by the coalition when the $i$-th NEMG is not included.

2) *Restrictive Condition*: a) Electrical power balance constraints

$$P_{mmg,t} = \sum_{k=1}^{N}(P_{k,t}^{PV} + P_{k,t}^{WT} - P_{k,t}^{Load} + P_{k,t}^{ev}), \quad (24)$$

$$P_{k,t}^{Load} = P_{k,t}^{NL} + P_{k,t}^{AL}, \quad (25)$$

where $P_{mmg,t}$ denotes the trading power of the MMG with the power distribution network at time $t$; $P_{k,t}^{LOAD}$ is the total user load power; $P_{k,t}^{NL}$ and $P_{k,t}^{AL}$ are the active power consumed by the i-th unresponsive and responsive loads in microgrid $k$ at time $t$, respectively.

b) Power distribution network interaction power constraints.

$$P_{line\_min} \leq P_{mmg,t}^{sell} \leq P_{line\_max}, \quad (26)$$

$$P_{line\_min} \leq P_{mmg,t}^{buy} \leq P_{line\_max}, \quad (27)$$

where $P_{line\_min}$ and $P_{line\_max}$ are the lower and upper limits of the interactive power of the power distribution network.

c) Trading power constraints between NEMGs.

$$P_{k,t}^{buy} = P_{k,t}^{sell}, \quad (28)$$

$$0 \leq P_{k,t}^{sell} \leq P_{k,t}^{max}, \quad (29)$$

$$0 \leq P_{k,t}^{buy} \leq P_{k,t}^{max}, \quad (30)$$

Where $P_{k,t}^{max}$ denotes the upper limit of power transfer between NEMGs.

B. *Lower-level optimization model*

Given their bidirectional energy storage capabilities, EVs are ideally positioned to enhance the efficient use and distribution of microgrid energy, as well as to facilitate electric energy feedback. On this basis, this study categorizes energy users and EV owners as distinct stakeholder groups within the demand side, and proposes a two-stage pricing mechanism in the lower level model, combining TOU electricity price and dynamic electricity price. TOU tariffs are designed to systematically guide users towards orderly electricity usage. Meanwhile, dynamic tariffs can be combined with TOU tariffs and the proportion of renewable energy remaining to set the tariffs for EVs. The two-stage pricing mechanism is shown in Fig. 5.

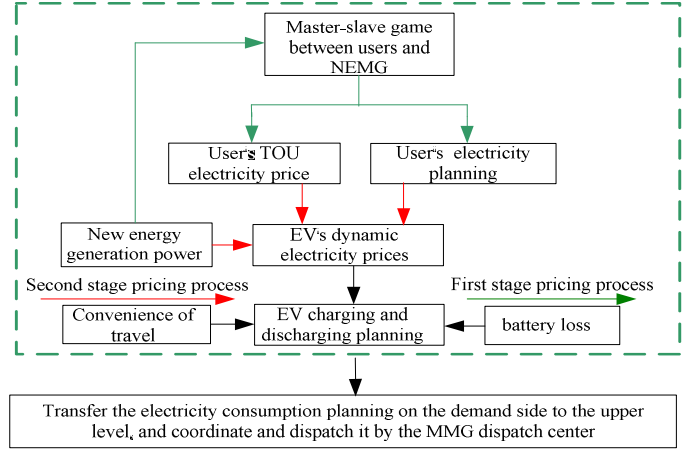

**Fig. 5.** Two-stage pricing flowchart.

The TPM is divided into two phases. First, the microgrid and the user, through the master–slave game, determine the user's power purchase price and power planning. Second, the microgrid management scheduling center, combined with the user's power price and power planning and other information, develops the EV electricity price. The owner of the car, according to the comprehensive satisfaction index, adjusts the charging and discharging planning, thus completing the graded formulation of the user and EV DR program. Finally, the results of the two-stage pricing mechanism will be transmitted to the upper model to achieve the joint scheduling of the MMG

1) *Objective Function*: In the first stage of the two-stage pricing mechanism, the microgrid operator, as the leader, issues user electricity price with (31) as an adaptation function. The users, as followers, respond to demand with (32) as a target, develop electricity consumption plans, and upload them to the NEMG dispatch side.

$$\max F_k = \sum_{t=1}^{24}\left(\left(\sum_{i=1}^{N_{NL}}C_{k,t}P_{k,i,t}^{NL} + \sum_{i=1}^{N_{AL}}C_{k,t}x_{i,t}P_{k,i,t}^{AL}\right) + \left(P_{k,t}^{ev}C_{k,t}^{ev}I_{k,t}^{char} + P_{k,t}^{ev}C_{k,t}^{ev}I_{k,t}^{dis}\right)\right) - F_{MMG,k}, \quad (31)$$

$$\min F_{L,k} = \sum_{t=1}^{24}\left(\sum_{i=1}^{N_{NL}}C_{k,t}P_{k,i,t}^{NL} + \sum_{i=1}^{N_{AL}}C_{k,t}x_{i,t}P_{k,i,t}^{AL}\right), \quad (32)$$

where $F_{MMG,k}$ and $F_k$ represent the operating cost as well as the revenue of microgrid $k$, respectively; $F_{L,k}$ represents the cost of electricity for users; $C_{k,t}$ and $C_{k,t}^{ev}$ are the transaction electricity price between microgrids and users as well as that between microgrids and EVs, respectively; $N_{NL}$ and $N_{AL}$ represent the total number of non-responsive loads as well as responsive loads; $P_{k,t}^{ev}$ is the charging and discharging power of the EVs; $I_{k,i,t}^{char}$ and $I_{k,i,t}^{dis}$ are the charging and discharging status variables of the EVs, where 1 represents charging and discharging respectively, and 0 represents not charging or discharging; $x_{i,t}$ represents the state of being able to respond to load participation in scheduling, 1 represents participation, and 0 represents non participation.

To fully quantify the impact of the user's DR on EV charging and discharging planning, so that it can further consume renewable energy and realize electric energy feedback, the dynamic tariff impact factor $R_{k,t}$ is introduced into the EV tariff formulation process, and the EV electricity price is solved by (33)–(35).

$$C_{k,t}^{ev} = C_{k,t} - R_{k,t} \tag{33}$$

The relevant parameters are as follows:

$$R_{k,t} = P_{k,t}^{UNP} / P_{k,t}^{LOAD}, \tag{34}$$

$$P_{k,t}^{UNP} = P_{k,t}^{WT} + P_{k,t}^{PV} - P_{k,t}^{LOAD}, \tag{35}$$

where $P_{k,t}^{UNP}$ is the unbalanced power of the microgrid; $R_{k,t}$ is the dynamic electricity price impact factor, which can reflect the impact of distributed energy output, user load, and other information on EV electricity prices.

To comprehensively consider the impact of battery loss and travel demand on EV owners' electricity consumption planning, this study selects maximizing the comprehensive satisfaction function as the objective function to optimize the EV electricity consumption strategy, as shown in (36):

$$\max \varsigma^{ev} = \frac{1}{N} \sum_{k=1}^{N} (\theta_k + \delta_k), \tag{36}$$

where $\theta_k$ and $\delta_k$ are the price satisfaction function and the ride satisfaction function accounting for battery losses, respectively, to assess the impact of battery losses due to charging and discharging and vehicle demand on electricity planning; $n$ represents the number of EVs participating in the dispatch.

Price satisfaction function considering battery depletion

$$\theta_k = 1 - \left| F_k^{EV} - F_k^{min} \right| / \left| F_k^{max} - F_k^{min} \right|, \tag{37}$$

Where $F_k^{max}$ and $F_k^{min}$ are the maximum and minimum expenses that EV owners can afford; $F_k^{EV}$ is the total cost of EV.

The relevant parameters are as follows:

$$F_k^{EV} = F_k^{ev} + F_k^{loss}, \tag{38}$$

$$F_k^{ev} = \sum_{j=1}^{n} \sum_{t=1}^{24} \left( P_{k,j,t}^{ev} C_{k,t}^{ev} I_{k,j,t}^{char} + P_{k,j,t}^{ev} C_{k,t}^{ev} I_{k,j,t}^{dis} \right), \tag{39}$$

$$F_k^{loss} = \sum_{j=1}^{n} \sum_{t=1}^{24} \left( \left| P_{k,i,t}^{ev} \right| \frac{C_{change}}{T_{max}^{ev}} \right), \tag{40}$$

where $F_k^{ev}$ and $F_k^{loss}$ are the EV charging and discharging costs and battery depletion costs, respectively; $C_{change}$ is the cost of replacing EV batteries; $T_{max}^{ev}$ is the total charge and discharge capacity of the EV battery at the end of its cycle life (capacity decay to 80% of the initial value).

Travel satisfaction function

$$\delta_k = \sum_{j=1}^{n} \left( 1 - \sum_{t=1}^{24} \left| P_{k,i,t}^{outmax} - P_{k,i,t}^{ev} \right| / \sum_{t=1}^{24} \left| P_{k,i,t}^{outmax} - P_{k,i,t}^{outmin} \right| \right), \tag{41}$$

where $P_{k,i,t}^{outmax}$ and $P_{k,i,t}^{outmin}$ are the power of EVs at the maximum and minimum travel satisfaction, respectively.

2) *Restrictive Condition*:

*a*) User electricity price constraints:

$$C_{k,t}^{min} \leq C_{k,t} \leq C_{k,t}^{max}, \tag{42}$$

where $C_{k,t}^{max}$ and $C_{k,t}^{min}$ are the upper and lower bounds of the microgrid's electricity prices for trading with users.

*b*) EV electricity price constraints:

$$0.7 C_{k,t} \leq C_{k,t}^{ev} \leq 1.3 C_{k,t}. \tag{43}$$

*c*) Response to changing load constraints:

$$P_{min}^{AL} \leq P_{k,t}^{AL} \leq P_{max}^{AL}, \tag{44}$$

where $P_{max}^{AL}$ and $P_{min}^{AL}$ are the maximum and minimum values of the responsive load change. In this study, the responsive load power change interval was set to 20% of the total load.

*d*) MMG system cost constraints:

$$F_{MMG} = \sum_{k=1}^{N} F_{MMG,k}. \tag{45}$$

*e*) EV charging and discharging power constraints:

$$P_{max}^{ev} \leq P_{k,i,t}^{ev} \leq P_{max}^{ev}. \tag{46}$$

*f*) EV charge and discharge state variable constraints:

$$0 \leq I_t^{dis} + I_t^{char} \leq 1. \tag{47}$$

*g*) EV charge state constraints:

$$SEV_{min} \leq SEV_t \leq SEV_{max}, \tag{48}$$

$$SEV_{t+1}E = SEV_t E + P_t^{ev} I_t^{char} \eta_{char} + \frac{P_t^{ev} I_t^{dis}}{\eta_{dis}}, \tag{49}$$

where $SEV_{max}$ and $SEV_{min}$ are the upper and lower limits of the EV charging state; $E$ is the rated capacity of the EV battery; $\eta_{char}$ and $\eta_{dis}$ are the EV charging and discharging efficiencies.

C. Model solving process

The model solving process is organized into a comprehensive iterative procedure, fundamentally segmented into two primary phases: initial setup and iterative optimization. Within these phases, the process involves multiple steps to refine the results, as illustrated in Fig. 6. Below is a detailed breakdown of each step:

**Phase 1: Initial Setup**

Step 1: The typical WT, PV and load output scenarios for a MMG system are obtained based on the measured WT and PV output data in a region and using the WT and PV uncertainty treatment methods cited in Section I.B.

Step 2: The parameters of MMG operators, users, and EVs are initialized. In this study, we set the number of populations $z$ to 40, the number of iterations e to 90, the population variation rate to 4%, the crossover probability to 80%, and the convergence error $\varepsilon = 0.01$.

**Phase 2: Iterative Optimization**

Step 3: Using the algorithm, the selling electricity prices of $m$ groups of NEMG operators are generated.

Step 4: $e = e + 1$.

Step 5: The user receives $m$ sets of selling prices, solves for the flexible electric load distribution using the CPLEX solver, calculates the current user cost of electricity, and returns the power purchase plan to the microgrid operator.

Step 6: The microgrid operator adjusts the EV electricity price through (33)–(35) and uses (36) as the objective function to obtain the EV power consumption planning in different time periods using the CPLEX solver.

Step 7: Based on the renewable energy generation data and the optimization results of the two-stage pricing mechanism, the power inequality of each NEMG is calculated and the relevant data is uploaded to the MMG dispatch center.

Step 8: The MMG dispatch center uses a solver to calculate the electricity trading volume between microgrids and between microgrids and the power distribution network, while retaining the current operating cost of the microgrid alliance.

Step 9: Using the Shapley method to allocate alliance costs, the operating cost $F_{mmg,k}$ of microgrid $k$ is obtained.

Step 10: Taking (31) as the fitness function, the genetic algorithm is utilized to generate the new selling price of electricity, and steps 4)–9) are repeated to calculate the profit $F_k$ of microgrid $k$, as well as to calculate the user's electricity cost $F_{L,k}$ according to (32).

Step 11: If $F_k^e > F_k^{e+1}$ then $F_k^{e+1} = F_k^e$, $F_{L,k}^{e+1} = F_{L,k}^e$; otherwise, $F_k^{e+1} = F_k^{e+1}$, $F_{L,k}^{e+1} = F_{L,k}^{e+1}$ 。

Step 12: If $|F_{L,k}^{e+1} - F_{L,k}^e| < \varepsilon$ and $|F_k^{e+1} - F_k^e| < \varepsilon$, then output the optimal scheduling policy and related parameters; otherwise, return to step 4.

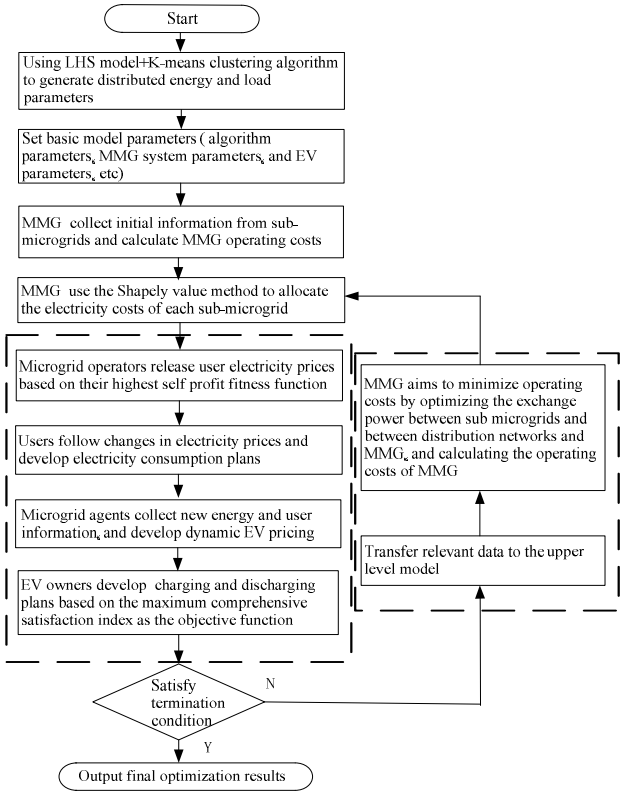

**Fig. 6.** Model solving flowchart.

IV. SIMULATION ANALYSIS

*A. Simulation system and settings*

The MMG system used in the simulation of this study consists of three NEMGs, and each sub-microgrid is connected to 130 EVs, of which 70 are controllable vehicles and 60 are uncontrollable vehicles. According to the travel pattern of EVs, the basic information of EVs such as connecting time to the grid, disconnecting time from the grid, and initial load condition can be obtained by using the Monte Carlo method [27,28]. The WT and PV output for the MMG system and basic load prediction curves are shown in Fig. 7. The relevant operating parameters of the MMG are shown in Table II.

Table II
System simulation parameters

| parameters | values | parameters | values |
| --- | --- | --- | --- |
| $C_{WT}$(yuan/kw) | 0.376 | $C_{PV}$(yuan/kw) | 0.428 |
| $\lambda_L$ (yuan/kw) | 0.17 | $v_N$(m/s) | 12 |
| $P_k^{max}$ /kw | 450 | $v_{ci}$(m/s) | 3 |
| $v_{co}$(m/s) | 25 | $P_{change}$ /yuan | 18 000 |
| $\eta_{char}$ | 0.95 | $\eta_{dis}$ | 0.95 |
| $P_{max}^{ev}$/kw | 3 | $T_{max}^{ev}$/kwh | 100000 |
| $SEV_{min}$ | 0.1 | $SEV_{max}$ | 1 |

The corresponding electricity price constraints are presented in Table III, and the MMG sells power to the power distribution network at 0.7 times the power purchase price.

Table III
Electricity Price Constraints at Different Times (Yuan/kWh)

| Time | User tariffs | Inter-microgrid tariffs | Power distribution trading tariffs |
| --- | --- | --- | --- |
| 1:00–08:00 | 0.8–1.0 | 1.1 | 1.2 |
| 09:00–11:00 | 0.9–1.2 | 1.3 | 1.4 |
| 12:00–20:00 | 1.2–1.4 | 1.5 | 1.65 |
| 21:00–0:00 | 1.0–1.3 | 1.4 | 1.5 |

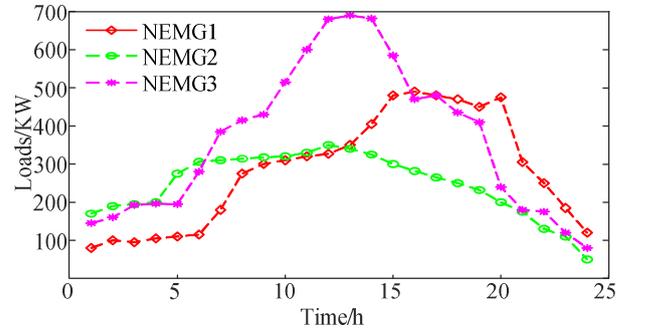

a) MMG's daily load prediction curve.

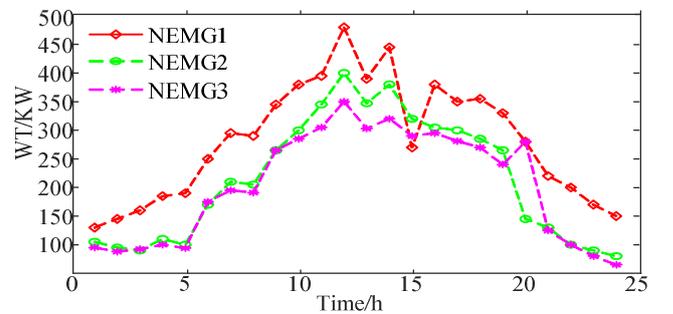

b) MMG's wind power prediction curve.

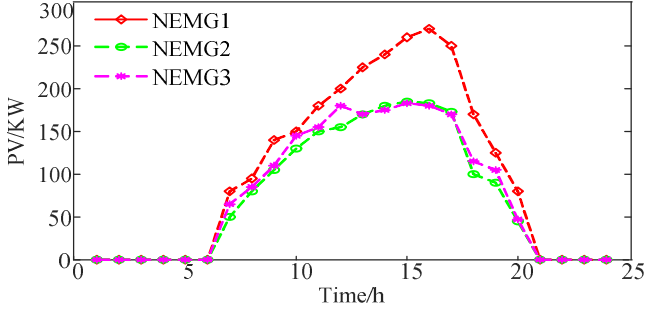

c) MMG's photovoltaic prediction curve.
**Fig. 7.** MMG's daily load, wind power, and photovoltaic prediction curve.

The simulation calculations in this study were performed using MATLAB R2018b software combined with the Yalmip plug-in to invoke the CPLEX solver, and the computer comprised an Intel Core i7 processor at 1.8 GHz.

### B. WP-PV-load output uncertainty treatment

We used a stochastic scenario simulation analysis to consider prediction errors. We used the treatment of uncertainty described in Section 2.C to generate 1000 wind power, PV, and loads scenarios using LHS and then reduced the generated scenarios using the K-Means clustering algorithm to obtain five typical joint power scenarios of WP, PV, and load with different probabilities. The probability distributions of the scenarios are shown in Table IV.

Table IV
Probability distribution of combined output scenarios for MMG

| Scenario 1 | Scenario 2 | Scenario 3 | Scenario 4 | Scenario 5 |
| --- | --- | --- | --- | --- |
| 15.99% | 20.73% | 30.23% | 14.51% | 18.54% |

Scenarios 6 and 7 were added to analyze the effect of uncertainty. Scenario 6: Multiply the obtained data of five typical combined WP, PV, and load output scenarios by the corresponding probabilities and then sum them as simulation data, that is, the method used in this study. Scenario 7 did not consider uncertainty; that is, it substituted the day-ahead forecast data into the model. The simulation results for each scenario are listed in Table V.

Table V
Comparison of results by scenario

| scenario | Gain/yuan | | | MMG running cost/yuan | MMG WP and PV abandonment rate/% |
| --- | --- | --- | --- | --- | --- |
| | NEMG 1 | NEMG2 | NEMG 3 | | |
| 1 | 13496.37 | 9365.48 | 11642.11 | 7454.63 | 12.37 |
| 2 | 13655.71 | 9530.67 | 11700.52 | 7347.57 | 11.56 |
| 3 | 13795.42 | 9552.93 | 11736.43 | 7394.36 | 11.04 |
| 4 | 13451.35 | 9311.25 | 11594.32 | 7429.85 | 13.06 |
| 5 | 13752.46 | 9468.14 | 11694.25 | 7443.24 | 12.27 |
| 6 | 13833.21 | 9522.57 | 11740.35 | 7469.51 | 10.66 |
| 7 | 13522.47 | 9341.36 | 11514.69 | 7480.34 | 13.51 |

The analysis shows that the difference in the operating cost benefits of the MMG system in the seven scenarios is significantly small. This is due to the power interconnection between the sub-microgrids to reduce the impact of output uncertainty, indicating that the prediction error has a small impact on the operating cost of the MMG system. By comparing Scenario 7 to Scenario 6, it can be seen that when the prediction error of the system is not taken into account, there is a large difference in the rate of WP and PV abandonment and operational benefits of the sub-microgrids; the rate of WP and PV abandonment increases by 2.85%. When errors were considered, Scenario 3 was close to Scenario 6 for all parameters, whereas Scenario 4 deviated slightly from Scenario 6 because Scenario 4 had a significantly smaller probability of occurrence than Scenario 3, with larger deviations in WP, PV, and load outputs for certain extreme scenarios. In summary, after considering the forecast error, the revenue of each main body increases slightly, and the WP and PV abandonment phenomena are effectively improved.

### C. Simulation system analysis

To verify the feasibility of the optimized scheduling scheme proposed in this paper, four schemes are set up as comparisons.

1) **Scheme** 1: This scheme involves electrical energy interactions between microgrid systems and adopts the two-stage pricing mechanism described in this article. It considers the DR behavior of different entities on the demand side.

2) **Scheme** 2: There is an electrical energy interaction between the NEMG systems in this scheme, but the customer base load and EVs are treated as a single entity on the demand side for DR, meaning the two-stage pricing mechanism proposed in this paper is not implemented on the demand side.

3) **Scheme** 3: While there is no power interaction between the microgrid systems in this scheme, the two-stage pricing mechanism proposed in this study is used on the demand side.

4) **Scheme** 4: This scheme features no collaboration between the multiple microgrids, and without considering DR behavior on the demand side, the microgrids sell electricity to loads at the same price as in Scheme 1.

The single solution time of the optimized scheduling method proposed in this paper was 5.58 s, and the total computation time was 501.79 s, which met the requirements of day-ahead scheduling. The optimization solutions were obtained for the above four schemes, and the results are listed in Table VI.

Table VI
Simulation results (Yuan)

| Schemes | NEMG | Operating income | Customers electricity costs | EVs electricity costs |
| --- | --- | --- | --- | --- |
| 1 | 1 | 13833.21 | 8284.76 | 657.36 |
|   | 2 | 9522.57 | 7139.54 | 787.25 |
|   | 3 | 11740.35 | 9844.32 | 913.57 |
|   | Total | 35096.31 | 25268.62 | 2358.18 |
| 2 | 1 | 12569.85 | 10469.78 | |
|   | 2 | 8511.53 | 9713.64 | |
|   | 3 | 10037.49 | 12394.82 | |
|   | Total | 31118.87 | 32578.32 | |
| 3 | 1 | 11226.42 | 9164.12 | 745.31 |
|   | 2 | 7709.61 | 8197.26 | 905.73 |
|   | 3 | 8447.35 | 11759.17 | 1007.21 |
|   | Total | 27383.38 | 29120.55 | 2658.25 |
| 4 | 1 | 8676.92 | 11832.17 | 1401.36 |
|   | 2 | 5030.71 | 9235.28 | 1471.62 |
|   | 3 | 6572.25 | 13039.85 | 1534.51 |
|   | Total | 20279.88 | 34107.30 | 4407.49 |

Compared to Scheme 2, Scheme 1 introduces a two-stage

pricing mechanism. After implementing this pricing strategy, the total electricity cost for different entities on the demand side of the MMG system—specifically, users and EVs—decreased by 15.20%, and the daily operating revenue of the MMG alliance correspondingly increased by 3977.44 yuan. This improvement is attributed to the fact that the two-stage pricing mechanism in Scenario 1 differentially treats users and EVs for collaborative scheduling, thus generating demand response (DR) activities that consider the sequential order of energy transactions. This DR behavior with coupled relationships effectively motivates EV owners to participate in the dispatch in a timely manner according to the surplus and shortage conditions of the NEMG. Consequently, it not only enhances the consumption of distributed energy but also improves the financial outcomes for all involved stakeholders [28]. For the MMG system, owing to the increase in the penetration rate of distributed energy within the microgrid, the electricity trading volume within the alliance is increased, thereby reducing the purchase of electricity from the power distribution network and increasing the overall economic efficiency of the alliance. Therefore, the two-stage pricing mechanism can be considered superior to the DR strategy that considers EVs and users as the same subject, and the strategy proposed in this paper has a synergistic and superimposed enhancement effect on the improvement of the electric energy interaction relationship of the MMG system.

Comparing Scheme 1 with Scheme 3, each sub-microgrid establishes a community of interest through the cooperative alliance mechanism, which creates a power exchange relationship between microgrids. The daily operating revenue of the MMG system rises by 28.17% based on the introduction of the two-stage pricing mechanism. At the same time, owing to the lower price of electricity traded between subnets compared to the trading price with the power distribution network, the electricity cost on the load side also decreased by 4352 yuan. Compared with Scheme 4, Scheme 3 reduces the electricity costs for users and EVs by 14.62% and 39.69%, respectively, proving that the proposed two-stage pricing mechanism has an improvement effect on the energy costs on the load side. Therefore, it can be considered that the two-stage pricing mechanism is effective and reasonable in guiding EVs to participate in scheduling and achieve a win–win situation for different subjects on the demand side.

From the analysis of the simulation results, it can be observed that the proposed two-stage pricing mechanism improves the interests of different stakeholders within the MMG system. The following section will further analyze the impacts of the scheduling strategy proposed in this paper on the MMG system incorporating EVs, so as to elaborate on the superiority of this scheduling strategy.

*D. Analysis of the impact of comprehensive scheduling strategies on users and EVs*

In the analysis of user DR, a master-slave game between the user and the NEMG is employed. Iterative calculations are conducted to achieve the Stackelberg equilibrium solution, ensuring that no participant could gain more by deviating from their assigned strategy. The outcomes related to the users' time-of-use electricity pricing and consumption plans post-demand response are detailed in Fig. 8.

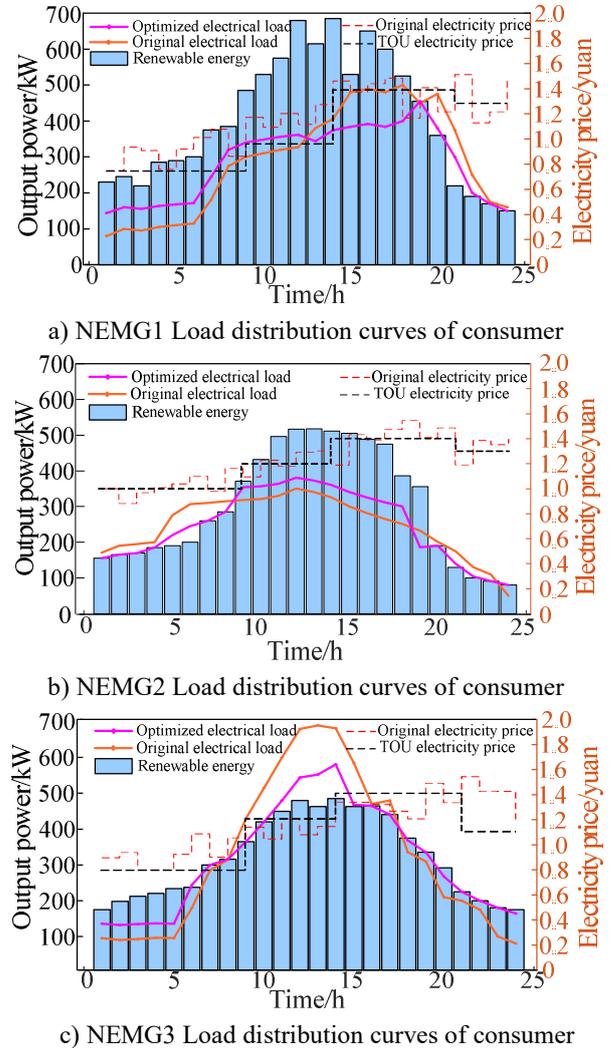

a) NEMG1 Load distribution curves of consumer

b) NEMG2 Load distribution curves of consumer

c) NEMG3 Load distribution curves of consumer

**Fig. 8.** Comparison of user electricity loads before and after the master–slave game.

As illustrated in Fig. 8, users in NEMG1 and NEMG3 have opted to increase their electricity consumption from 0:00-7:00 and decrease it from 18:00-24:00. In contrast, most users in NEMG2 reduce their consumption from 0:00-8:00. This behavior is consistent with the time-of-day tariffs optimized by the microgrid operator, encouraging customers to use more electricity during low-price periods and less during high-price periods. Additionally, the optimized user load curves more closely align with the renewable energy generation curves, effectively reducing the peak-to-valley differences in user load and thereby validating the accuracy of the results. The tariffs for electric vehicles (EVs) and their power consumption plans for each time period are determined using Eqs. (33) and (37) respectively, and are presented in Fig. 9.

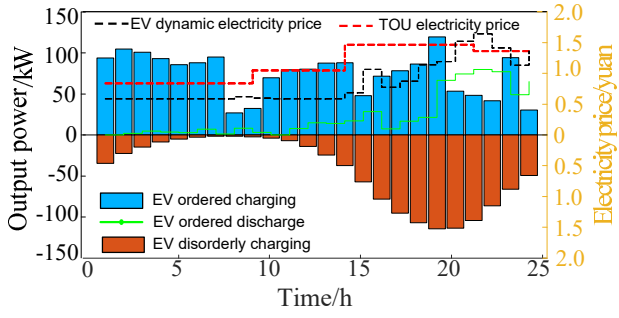

a) NEMG1: EV load distribution curves

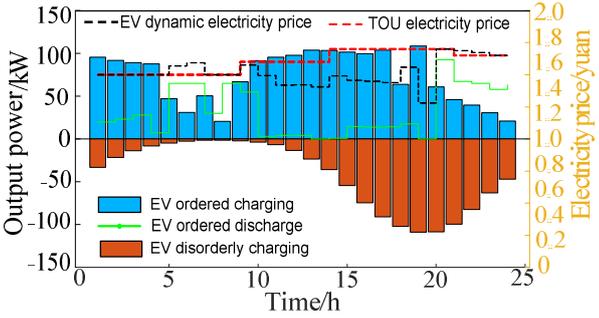

b) NEMG2: EV load distribution curves

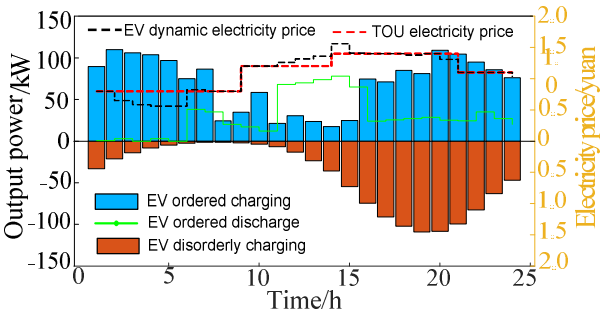

c) NEMG3: EV load distribution curves

**Fig. 9.** Load distribution curves for EVs across three NEMGs under different pricing strategies

As illustrated in Fig. 9, orderly charging/discharging of electric vehicles is defined as the behavior of car owners adjusting their charging/discharging activities in response to dynamic electricity price changes set by the TPM. Conversely, disorderly charging describes scenarios where electric vehicles are treated as loads without price guidance, leading to suboptimal electricity consumption planning. It is observed that without tariff guidance, most EV loads peak during the 17:00-22:00 time period. During these peak hours, the microgrid often cannot meet the EV charging demand, necessitating the purchase of electricity from the distribution network and subsequently increasing the operating costs of the system. When incentivized by a dynamic time-of-use tariff, which lowers rates during periods of power surplus in the microgrid, EV owners are encouraged to shift their charging from the 16:00-24:00 peak to these off-peak periods. This shift effectively alleviates the microgrid's power consumption pressure during peak hours. Additionally, EVs can contribute to grid stability and generate profit by selling back electricity during periods of high electricity prices.

To further demonstrate the effectiveness of EV participation in load management, the power unevenness measurement curve is depicted in Fig. 10.

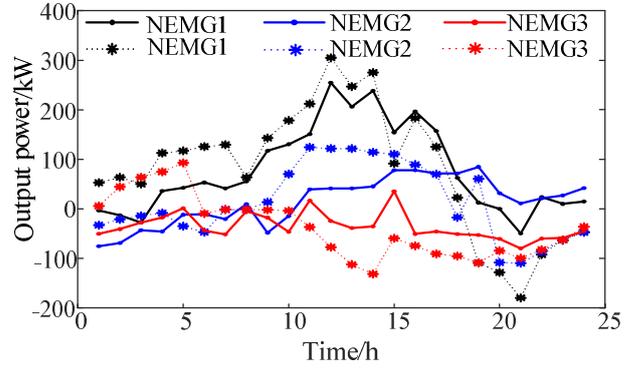

**Fig. 10.** Power imbalance of the microgrid before and after the participation of EVs in scheduling.

The dashed and solid lines in Fig. 10 represent the power inequality curves of the EVs when they are not involved in scheduling and when they are actively participating, respectively. It is apparent that the power inequality curve for the microgrid is less volatile and smoother when EVs engage in orderly charging and discharging. This indicates that the EV tariffs formulated by the two-stage pricing mechanism fully reflect the changes in residual power within the NEMG, and the dynamic tariffs they formulate can effectively guide EV owners to further participate in charging and discharging based on the consideration of the user demand response, proving the effectiveness of the pricing strategy in promoting the participation of EVs in dispatch. The amounts of power purchased and sold to the distribution network before and after the adoption of the TPM, as well as the amount of renewable energy consumption, are shown in Fig. 11 and Fig. 12.

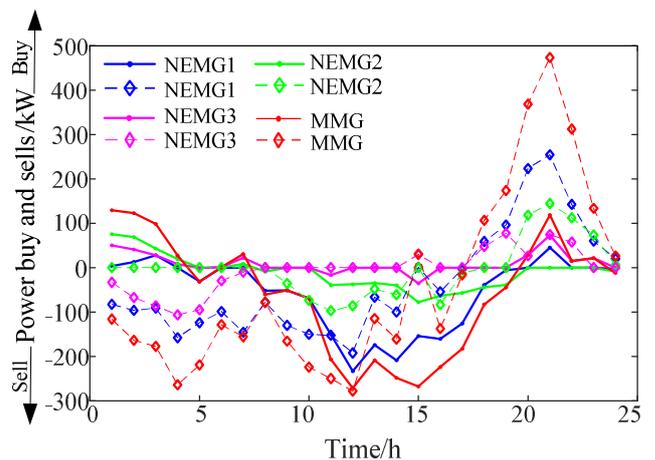

**Fig. 11.** Comparative analysis of electricity transactions with the distribution network pre- and post-TPM

In Fig.11, the dashed and solid lines represent the electricity purchased/sold by the system to the distribution system. It is observable that the sales of electricity during the period of 0:00-12:00 and the purchase of electricity during the period of 17:00-24:00 have shown a downward trend compared to before adopting this strategy. This trend is attributed to the fact that the strategy proposed in this paper can hierarchically guide the participation of users and EVs in scheduling, shifting the peak loads at other times of the day, thus allowing more power to be consumed by the system. Furthermore, between 12:00-17:00, electricity sales by the multi-microgrid system have increased compared to scenarios without this strategy. This rise in sales corresponds with peak PV power generation during these hours and is further supported by the reduction in peak-to-valley load differences, a result of the optimal dispatch strategy. The changes in the total amount of electricity purchased/sold by MMG are detailed in Table VII.

Table VII
The total amount of electricity purchased/sold by the system to the distribution network (kW)

| System | NEMG1 | NEMG2 | NEMG3 | MMG |
|---|---|---|---|---|
| Adopting TPM | 155.27/ 1484.91 | 304.79/ 405.93 | 468.86/ 53.80 | 928.92/ 1944.64 |
| Not using TPM | 855.14/ 1732.58 | 455.80/ 505.84 | 311.18/ 429.00 | 1622.12/ 2667.42 |

From Table VII, it is evident that with the adoption of the TPM, the total electricity purchased and sold by the MMG to the distribution network decreases by 693.20 kW and 722.78 kW, respectively. These reductions, corresponding to 42.73% and 27.10%, allow for increased internal consumption of electricity by the MMG.

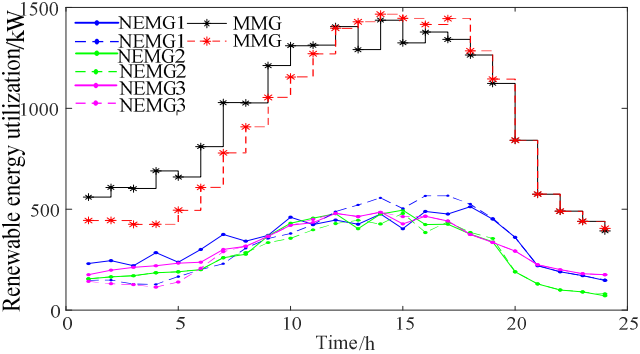

**Fig. 12.** Distributed energy consumption before and after the adoption of the TPM

In Fig. 12, the dashed and solid lines represent the electricity purchased/sold by the system to the distribution network and the amount of distributed energy consumed before and after the adoption of the TPM, respectively. It is observed that with the introduction of the TPM, the consumption of renewable energy in the MMG system has significantly increased, from 20,121.22 kW to 22,160.06 kW. This outcome aligns with the conclusion that the TPM effectively guides users and EVs to participate hierarchically in scheduling, thereby promoting distributed energy consumption. Thus, it confirms the TPM's effectiveness in promoting more efficient energy use within the MMG system.

*E. Analysis of the impact of comprehensive scheduling strategies on the MMG*

From the above analysis, it can be observed that the MMG cooperative alliance can improve the energy supply economy compared to the NEMG by absorbing the internal excess power through the power interaction between sub-microgrids. Therefore, if the tradable amount of electricity between sub-microgrids can be increased through the scheduling strategy, the energy supply economy of the MMG system can be improved again on the basis of the MMG cooperative alliance model. To illustrate these dynamics more concretely, a comparison of the electricity trading volumes within the MMG system is shown in Fig. 13.

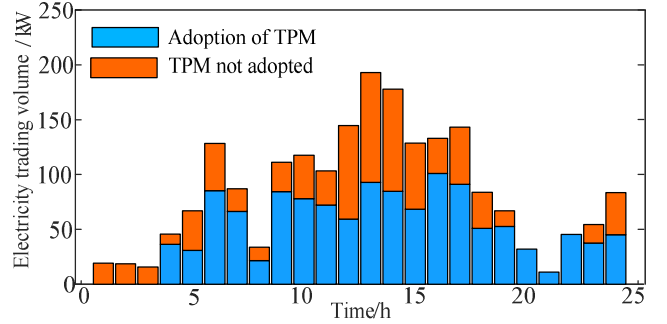

**Fig. 13.** Comparison of electricity trading volume in the MMG system.

As observed in Fig. 13, the introduction of the two-stage pricing mechanism significantly increases the volume of electric energy transactions within the MMG. This increase is primarily due to the DR behavior encouraged by the pricing mechanism, which synergistically enhances the electric energy interactions within the MMG system. As a result, the TPM fosters more flexible energy interactions and boosts electricity trading volumes between sub-microgrids. An in-depth analysis follows, examining the intrinsic reasons for this synergistic and cumulative enhancement attributed to the TPM.

The first stage of the two-stage pricing mechanism aligns user DR more closely with the patterns of new-energy generation. This alignment helps mitigate instances where the MMG cannot meet sudden increases in electricity demand due to load surges, facilitating smoother electricity trading among NEMGs. Thus, user DR is a key factor in enhancing the overall economics of the MMG system, leveraging the advantages of a cooperative MMG alliance.

The second stage expands the coalition's electricity trading volume by directing EVs to participate in the scheduling, building on the foundation of user DR. During the period of power shortage, expands the coalition's electricity trading volume by directing EVs to participate in the scheduling, building on the foundation of user DR. Conversely, during periods of electricity surplus, EVs are incentivized to charge at lower tariffs. The stored electricity can be utilized during power troughs; although primarily fed back into their own microgrid, it can also serve other users and EVs through the alliance. This capability significantly increases the alliance's traded electricity volume, as it enables consumption by all participating sub-microgrids.

*F. Comparison of different scheduling strategies*

To validate the superiority of the proposed scheduling strategy (Method 1), we conducted a comparative analysis against three alternative scheduling methods. Notably, as the current application scenarios for integrated scheduling strategies that include EVs predominantly involve single-microgrid systems, we applied Methods 3 and 4 when only one NEMG is present in an independently operating MMG system. The comparison methods are outlined as follows:

**Method 2**: Consumers and EVs on the load side are regarded as a single entity, and the load side of the sub-microgrid performs a demand response with the objective of minimizing the load peak-to-valley mean-square deviation and load-side cost of electricity.

**Method 3**: Customer information is used as a known quantity to set EV tariffs based on customer tariffs and respond to EVs with the objective of minimizing the cost of EV electricity consumption and the mean squared deviation of transactions with the distribution network.

**Method 4**: Scheduling EVs with the objective function of maximizing EV travel satisfaction and minimizing the cost of electricity consumption without considering the influence of users on EVs.

The optimization solutions were obtained for the above four schemes, and the results are listed in Table VIII.

Table VIII

Comparison of different strategies

| Methods | Load-side electricity costs/yuan | System Operating Costs/yuan | Purchase/sale of electricity to distribution networks/kW | WP / PV abandonment rate |
|---|---|---|---|---|
| 1 | 27626.80 | 7554.63 | 592.87/1964.47 | 10.56% |
| 2 | 30076.24 | 8721.82 | 718.21/2164.39 | 11.73% |
| 3 | 9876.33 | 3571.26 | 392.87/764.47 | 12.94% |
| 4 | 10172.68 | 3735.51 | 454.75/893.19 | 13.31% |

From the above table, it can be seen that the load-side electricity cost and system operation cost of Method 2 increased by 8.87% and 15.45%, respectively, and the rate of WP and PV abandonment increased by 1.17% compared with that of Method 1. The cost of electricity consumption and system operation cost of Strategy II increased by 8.87% and 15.45%, respectively. This is because Method 2 integrates EVs into the customer load for uniform scheduling, ignoring the subjective initiative of EVs when they are connected in large quantities, and fails to fully engage EVs in the V2G process, consequently leading to suboptimal economic performance and inefficient utilization of distributed energy resources.

The cost of load-side electricity consumption increases by 934.21 yuan and 1,230.56 yuan for Methods 3 and 4, respectively, compared to that in method 1 (NEMG 1). Moreover, parameters such as system operating costs and purchased and sold electricity also increase. On the one hand, this result is due to the fact that strategies III and IV are unable to obtain cheap electricity through power interactions between microgrids; on the other hand, these two strategies ignore the coupling relationship between users and EVs when they participate in demand response. Comparing Method 3 and Method 4, it can be seen that when Method 3 adjusts the four tariffs according to the load fluctuation, its demand response scheme will be more accurate, and thus, the optimization result is better than that of scheme IV, which also verifies the superiority of the strategy proposed in this paper.

In summary, the integrated scheduling strategy introduced in this study outperforms current strategies by effectively utilizing the synergistic incentives of TOU tariffs and dynamic tariffs. It comprehensively considers the coupling relationships between various stakeholders on the load side, guiding the participation of users and electric vehicles in demand response more accurately. This approach not only enhances the economic efficiency of the multi-microgrid system but also optimizes the involvement of different interest groups on the demand side.

### G. Assessment of the Scalability of Proposed Approach

To evaluate the scalability of the proposed approach, we analyzed the solution times for the MMG system across different scales within the same simulation environment. The results are summarized in Table IX.

Table IX

Comparison of simulation computation time

| Installed capacity of WT and PV/MW | Number of EVs | simulation time /s | | |
|---|---|---|---|---|
| | | TPM calculation time | Single iteration time | Total time |
| 1.5/1 | 390 | 4.39 | 5.58 | 501.79 |
| 2/1.5 | 510 | 5.45 | 6.73 | 606.04 |
| 3.9/2.5 | 1020 | 8.62 | 10.15 | 913.71 |

From Table IX, it is clear that as the number of EVs in the MMG system increases by 30.76% and 161.54%, the total computation time increases by 20.78% and 82.09%, respectively. This trend demonstrates a gradual escalation in computation time in response to the rapid expansion of the system size. Importantly, all computation times are well below 1 hour, confirming that the system is capable of handling scheduling on an hourly basis efficiently. This efficiency is largely due to the TPM, which simplifies the multi-objective optimization challenge into two nonlinear planning problems during the lower model solving process. This simplification significantly reduces the computational complexity of the system's optimal scheduling strategy. Furthermore, the cooperative alliance strategy at the MMG system level integrates all subnets into a single entity, thereby reducing the need for frequent electricity transactions between individual microgrids and the distribution network. Such integration not only lowers the computational burden but also shortens the total computation time.

To further explore scalability, this work analyzes how increasing the number of sub-microgrids in the MMG affects optimization solution speeds. Accordingly, Fig. 14 illustrates the optimization solution times for varying numbers of microgrids using the proposed approach.

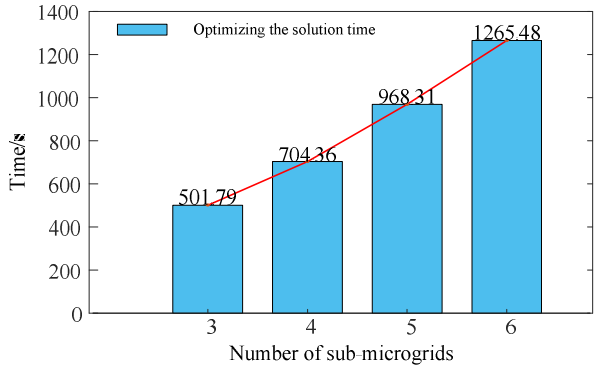

**Fig. 14.** Optimize scheduling time distribution with different numbers of sub-microgrids.

As shown in Fig. 14, as the number of sub-microgrids increases, the optimization solution time of the model will increase significantly, but the computational solution time still meets the requirements of day-ahead scheduling. This is due to the fact that, when the number of sub-microgrids increases, the interconnection and synergy of the system becomes more complex, which increases the difficulty of optimization search, and the increase in the amount of computation and the decrease in the efficiency of the algorithms will also become an important factor in restricting the computation time of the strategy optimization solution. Therefore, in order to alleviate the problems caused by the increase in the number of microgrids on the optimal scheduling strategy solving for MMG, an in-depth analysis of the more complex network structure of multi-microgrid systems is needed, as well as more advanced optimization algorithms to improve the optimization solving efficiency.

## V. CONCLUSION

This paper presents an optimal dispatch strategy for EVs and MMG systems, grounded in the concept of a cooperative alliance. This strategy effectively integrates resources from multi-level stakeholders within the MMG system, yielding significant economic benefits for all involved parties. The key conclusions drawn from this study are as follows:

1) The MMG cooperative alliance strategy developed in this research maximizes the potential of each sub-microgrid and significantly enhances their operational flexibility. Through the strategic utilization of electric energy across microgrids, the daily operating revenue of the MMG system is substantially increased. This approach ensures mutual benefits and fosters a win-win scenario for all participating microgrids.

2) The implementation of a two-stage pricing mechanism within the MMG system has effectively moderated the electric loads of users and optimized the charging and discharging cycles of EVs. This adjustment makes energy consumption more efficient and balanced.

3) Simulation analyses confirm that the integrated scheduling strategy delineated in this paper effectively coordinates the diverse interests within the MMG system, particularly with the inclusion of EVs. This approach facilitates dynamic electric energy flow and enhances the energy support capabilities across different NEMGs within the system.

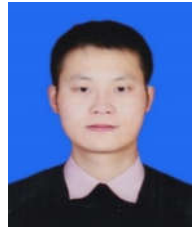

**Yang Li** (S'13–M'14–SM'18) was born in Nanyang, China. He received his Ph.D. degree in Electrical Engineering from North China Electric Power University (NCEPU), Beijing, China, in 2014.
Currently, he is a professor at the School of Electrical Engineering, Northeast Electric Power University, Jilin, China. From Jan. 2017 to Feb. 2019, he was a postdoc with Argonne National Laboratory, Lemont, United States. His research interests include sustainable energy, AI-driven power system stability analysis, smart grid, and integrated energy system. He is featured in Stanford University's List of the World's Top 2% Scientists and Elsevier's Most Cited Chinese Researchers for the years 2022-2023. He serves as an Associate Editor for several prestigious journals, including IEEE Transactions on Sustainable Energy, IEEE Transactions on Industrial Informatics, and IEEE Transactions on Industry Applications. He is also a Young Editorial Board Member of Applied Energy.

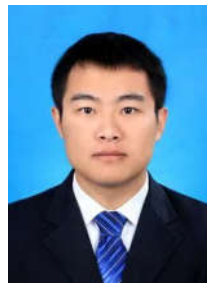

**Hengyu Zhou** was born in Liaoning, China, in 1994. He received the B.E. degree in 2016 from Liaoning Shihua University, Fushun, China. He received the M.E. degree in 2020 from Northeast Electric Power University Jilin, China.
He is currently working in State Grid Yingkou Electric Power Supply Company, Yingkou, China. His research interests include distributed energy storage configuration in distribution network.

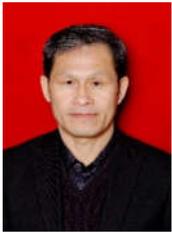

**Yonghui Nie** (M'17) received the M. Eng. degree in electrical engineering from Northeast Electric Power University, Jilin, China, in 2006 and the Ph. D. in electrical engineering from Xi'an Jiaotong University, Xi'an, China, in 2014.
He has been a Professor at Northeast Electric Power University, Jilin, China, since 2013. His current research interests include optimal operation, stability analysis, and control of power systems.

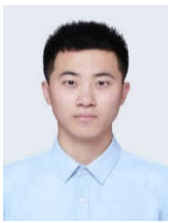

**Zhi Li** received the B.S. degree in electrical engineering from Qilu University of Technology, Shandong, China, in 2022. He is currently pursuing a master's degree at the Northeast Electric Power University, China. His main research interests are power system stability analysis and control.

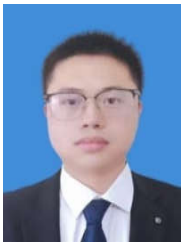

**Jie Zhang** received the master's degree in electrical engineering from Northeast Electric Power University, China, in 2024.
He has been an engineer at State Grid Chengwu Electric Power Supply Company. His research interests include the optimal operation, stability analysis, and control of power systems.

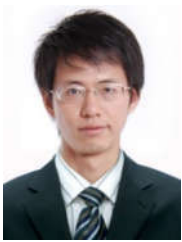

**Lei Gao** was born in Shandong Province, China in 1980. He received his PhD degree from Shandong University, China in 2014, and M.S degree from Northeast Electric Power University, China in 2006.
He is currently a professor of China Electric Power Research Institute (CEPRI). His main research interest includes power system analysis and control.